\documentclass[a4paper,fleqn]{cas-dc}
\usepackage[authoryear,longnamesfirst]{natbib}
\title{Design of laser uniform illumination system based on aspheric lens and compound ellipsoidal cavity}
\author{Yu Lu$^{\ast}$\thanks{*Corresponding author}}
\author{Xiangxiang Zhang}
\author{Ruilong Wu}
\author{Yi Yang}
\tnotetext[cor1]{Corresponding author  :E-mail address:luyu@tute.edu.cn}
\address{School of Optoelectronics Information science and Engineering, Tianjin University of Technology and Education, Tianjin 300350, China}
\def\tsc#1{\csdef{#1}{\textsc{\lowercase{#1}}\xspace}}
\tsc{WGM}
\tsc{QE}
\newpage
\usepackage{graphicx}
\usepackage{float}
\usepackage{subfigure}
\usepackage{booktabs}
\makeatletter
\renewcommand{\maketag@@@}[1]{\hbox{\m@th\normalsize\normalfont#1}}
\makeatother
\usepackage{natbib}
\setcitestyle{numbers,square}
\begin{abstract}
	In order to achieve uniform laser illumination with small aperture diameter and large field Angle,study laser active illumination system.An aspheric mirror combined with a composite ellipsoidal cavity is designed to achieve uniform illumination in this paper.Through an aspheric mirror,the fundamental mode of Gaussian beam is shaped into double Gaussian radiation and Flat-top radiation.The double Gaussian radiation rays are reflected again by the complex ellipsoidal cavity and decomposed into equal radiation flux,which is superimposed with the through Flat-top radiation rays to form a uniform distribution.The parameters of the complex ellipsoidal cavity are obtained by mapping equalization algorithm.After the superposition of the aspherical transmission Flat-top shaping and the composite ellipsoidal cavity secondary reflection shaping,the aperture is 29.7mm,whose aperture angle is 84.0 degrees,and the uniformity is 92.7$\%$ with 2m distance and 3.6m diameter.The optimization of uniformity is influenced by three factors:$RMS$,transmission and reflection power density ratio $M_{T/R}$ and transmission and reflection overlap degree.$RMS$ and $M_{T/R}$ determine the design effect of the composite ellipsoidal cavity, which depends on the maximum reflection Angle and transmission Angle.$M_{T/R}$ is negatively correlated with the maximum reflection of Angle,and $RMS$ is positively correlated with the transmission Angle.When the maximum reflection Angle is set to 32.0 degrees and the transmission Angle to 8.0 degrees,the minimum root-mean-square focusing radius is 108.6um,and the minimum effective transmission reflection power density ratio is 1.07.The degree overlap of transmission and reflection directly affects the uniformity of the target plane.The degree of transmission and reflection is adjusted by setting an adjustment factor.When the adjustment factor is 0.9,the uniformity of the target plane reaches the maximum.
\end{abstract}

\begin{document}
		\let\WriteBookmarks\relax
		\def\floatpagepagefraction
		\def\textpagefraction{.001}
		\let\printorcid\relax
		\maketitle
		\section{Introduction}
		Optical imaging system can be divided into passive imaging and active imaging. Owing to the weak return signal, background light interference and other factors, and the traditional passive imaging cannot provide sufficient contrast and resolution, thus traditional passive measurement imaging system cannot effectively identify the target \cite{bib:1}.According to research reports, near infrared light source can effectively suppress background interference, among which 808nm semiconductor laser source with mature technology has a long transmission distance in the air and can effectively penetrate fog and rain environment\cite{bib:2}.Therefore, the laser active lighting system with excellent directionality and high brightness can improve the effective imaging ability in dark environment\cite{bib:3,bib:7}.
		\par At present, the research on laser active lighting imaging system is attached to great importance in the domestic and overseas. OBZERV Night Vision Technology Company of Canada has developed a variety of lighting imaging systems, among which ARGC-750 and ARGC-2400 DALIS can effectively identify targets within 15km\cite{bib:8}. INTEVAC has designed the LIVAR4000 remote laser lighting gating imaging system with a maximum illumination of 15mrad\cite{bib:9}. The portable gas lighting imaging system designed by KJ Nutt can effectively identify methane gas monitoring within a range of 0.3-0.5m at a distance of 3m\cite{bib:10}. The laser active lighting system developed by Liu Tao et al. achieves a circular lighting area with a diameter of 10m at a distance of 200m\cite{bib:11}. The laser lighting system developed by Haijing Zheng et al can achieve uniform illumination with a diameter of 1.8m at a distance of 120m\cite{bib:12}.\par Studies on laser active imaging system have found that uneven target illuminance and large dynamic range of illuminance are key factors affecting imaging quality\cite{bib:13,bib:14}. Uniform lighting system can effectively eliminate background interference, extract targets from complex backgrounds, improve imaging quality, reduce image processing difficulty, and improve information extraction speed and imaging quality\cite{bib:15}. Uniform lighting system includes short and long - range lighting. Short-range lighting mainly adopts the method of microlens array, free-form surface transreflection element, etc., to achieve short-range and small-field uniform lighting\cite{bib:16,bib:23}. The lens group method and fiber waveguide method can realize the long distance and small field of view uniform illumination\cite{bib:24,bib:26}. Aspherical lenses are often used for uniform laser shaping, but a set of aspherical lenses can only homogenize a laser beam with a fixed light energy distribution, so this method is not universal and difficult to manufacture\cite{bib:27}. Now the large field of view illumination uses LED array structure mostly to obtain large field of view illumination\cite{bib:28}. The realization of small aperture and large field of view uniform laser lighting system will expand the application of laser in daily life, industry and military and other fields, such as security night vision lighting, laser projection lighting, smart city lighting, gas identification lighting, machine vision 3D imaging and ballistic path monitoring, etc. \cite{bib:29,bib:31}.
		\par In this paper, we use 808nm laser source with a beam waist radius of 5mm Gaussian distribution, and propose a method that combines aspherical lens and compound ellipsoid cavity to achieve uniform illumination with small aperture and large field of view. First, the laser is shaped into a 32°Flat-top double Gaussian distribution at the target plane after passing through an aspheric lens with a focal length of 8mm, where 0$ \sim $8°is flat-top power density distribution and 8$ \sim $32°is double-Gaussian power density distribution. Then, the composite ellipsoidal cavity is designed with the common first focus of the aspherical lens focusing spot, and the parameters of the composite ellipsoidal cavity are optimized by introducing an adjustment factor, and the Gaussian distribution is shaped into a flat-top distribution in the target plane. The target surface 2m away from the optical system forms a uniform radiation distribution with a radius of 1.8m and a uniformity of 92.7$\%$. The field of view angle of the system reaches 84.0°and the energy utilization rate is 97.8$\%$. The uniform laser lighting system has a simple design structure and wide applicability, which can satisfy different field of view lighting, and lays a foundation for laser lighting target monitoring in dark vision environment.
		\section{Uniform lighting system design}
		For it is difficult to achieve uniform radiation output of a Gaussian beam with a large field of view in only transmission mode, this design uses an aspherical lens combined with a composite ellipsoid cavity to achieve radiation distribution of a small aperture and a large field of view. And this design is composed of transmitted radiation and reflected radiation distribution. The schematic diagram is shown in Fig.1.
		\begin{figure}[htbp]
			\centering
			\includegraphics [scale=0.3]{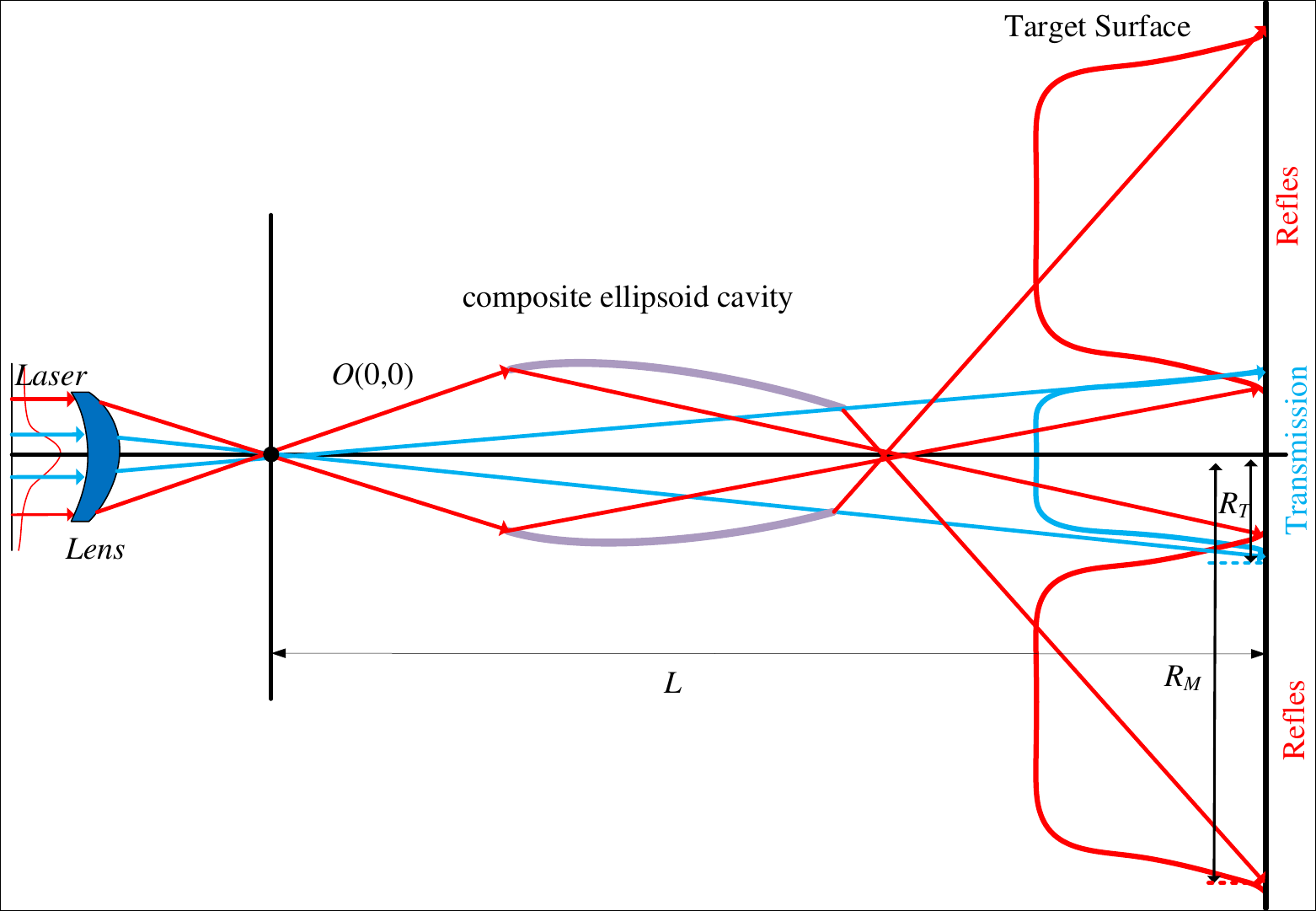}
			\caption{Overall structure of the system}\label{fig1}
		\end{figure}
		\subsection{Aspherical lens design}
		\subsubsection{Target plane light intensity distribution model}
		The aspherical surface shapes the Gaussian beam into flat-topped and double-Gaussian distributions, and achieves the divergence angle required by the reflection of the composite ellipsoidal cavity, as shown in Fig. 2. In the Fig. 2,  $\theta_{T}$ and $\theta_{m}$ are set as the transmission and reflection divergence half angles of the composite ellipsoid cavity, L is the distance from the focus to the target plane, and the waist radius before and after Gaussian distribution shaping is $\omega_{1}$ and $\omega_{2}$, respectively. Because of the symmetry of the structure , only the upper semiaxial distribution of the outgoing rays is analyzed below.
		\begin{figure}[htbp]
			\centering
			\includegraphics [scale=0.5]{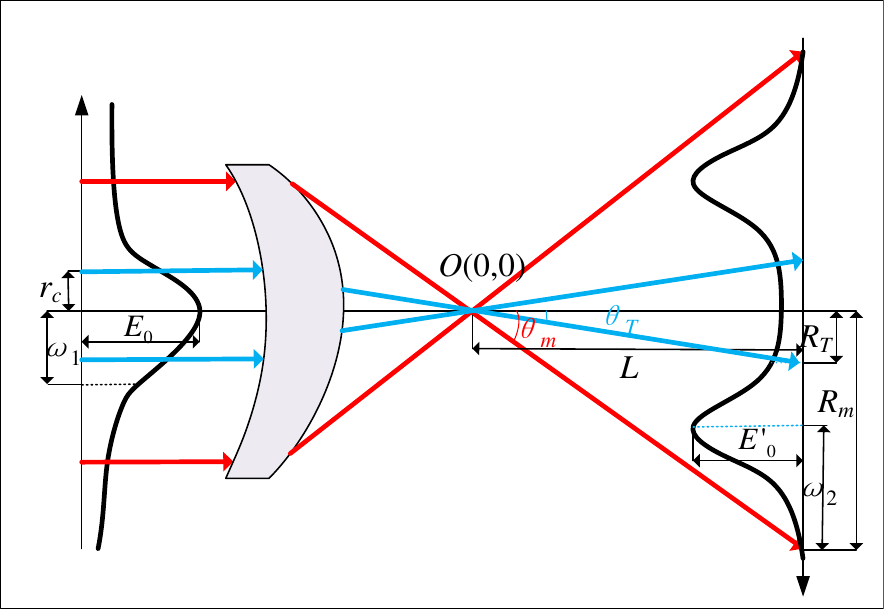}
			\caption{Schematic diagram of light intensity distribution before and after aspherical lens shaping}\label{fig1}
		\end{figure}
		\par It can be seen from Fig 2 that $\theta_T{\sim}\theta_m$ energy is a Gaussian distribution, and $0{\sim}\theta_T$ energy is a rectangular distribution, and its distribution function is shown in Formula(1).
		\begin{small} 
			\begin{equation} 
				E_{out}\left(R\right)=\left\{\begin{aligned}&E_{0}^{\prime}\exp\left[-\frac{\left(R_{T}-R_{m}+\omega_{2}\right)^{2}}{\alpha_{2}^{2}}\right]&&0\leq R\leq R_{T}\\&E_{0}^{\prime}\exp\left[-\frac{\left(R-R_{m}+\omega_{2}\right)^{2}}{\alpha_{2}^{2}}\right]&&R_{T}<R\leq R_{m}\end{aligned}\right.
			\end{equation} 
		\end{small}
		 \par $E_{0}$ and $\mathrm{E}_0^{\prime}$ are the incident and emergent rays peak power densities, respectively, $R_{T}=L\mathrm{tan}\theta_{T},R_{m}=L\mathrm{tan}\theta_{m}.$
		\subsubsection{Aspherical lens design method}
		The radiant flux density of single-mode laser source is Gaussian distribution after beam expansion.The formula is $E_{in}\left(r\right)=E_{0}\exp\left(-\frac{2r^{2}}{\omega_{1}^{2}}\right)$.The radiant flux distribution is $\Phi_{in}(r)=\frac{\pi E_{0}\omega_{1}}{2}\Bigg(1-\exp\Bigg(-\frac{2r^{2}}{\omega_{1}^{2}}\Bigg)\Bigg)$, The radiant flux distribution after reshaping can be expressed as Formula(2) by Formula(1).
		\begin{small} 
			\begin{equation} 
				\Phi_{out}(R)=\left\{\begin{aligned}&\pi E_0^{\prime}R^2\exp\left[-\frac{2(R_T-R_m+\omega_2)^2}{\omega_2^2}\right]\;0\leq R\leq R_T\\&2\pi\int_{R_T}^RRE_0^{\prime}\exp\left[-\frac{2(R-R_m+\omega_2)^2}{\omega_2^2}\right]dR\;R_T<R\leq R_m\end{aligned}\right.
			\end{equation} 
		\end{small}
		\par According to the law of conservation of energy, that is, $\Phi_{in}(r){=}\Phi_{out}(R)$, which can be expressed as $f(r,R_{T})=0$, the $R(r)$ mapping relationship can be calculated to establish the evaluation function to optimize the surface parameters of the aspherical lens. $f(r,R_{T})=0$ is given by Formula(3).
		\begin{figure*}[htbp] 
			\begin{equation}
				f(r,R)=\begin{cases}\frac{\pi E_0\omega_1}{2}\bigg(1-\exp\left(-\frac{2r^2}{{\omega_1}^2}\right)\bigg)-\pi E_0^{\prime}R^2\exp\bigg[-\frac{2(R_T-R_m+\omega_2)^2}{{\omega_2}^2}\bigg]&0\le r\le r_T\\\frac{\pi E_0\omega_1}{2}\bigg(1-\exp\left(-\frac{2r^2}{{\omega_1}^2}\right)\bigg)-2\pi\int_{R_T}^{R}RE_0^{\prime}\exp\bigg[-\frac{2(R-R_m+\omega_2)^2}{{\omega_2}^2}\bigg]dR&r_T\le r\end{cases}
			\end{equation}
		\end{figure*}
		\par According to the Formula(3), $r_T$ is the boundary ray with a divergence half Angle of $\theta_{T}$ after the aspherical lens shaping, corresponding to the target plane $R_T$, which can be seen from Fig.2 that the Gaussian waist radius $\omega_{2}$ after reshaping depends on the transmission and reflected radiation power distribution. In order to achieve uniform distribution of radiation flux density on the target plane, the ratio of transmitted radiation flux $\Phi_{T}$ to reflected radiation flux $\Phi_{T}$ is equal to the ratio of corresponding radiation area according to the energy of the reshaped light and the area of the mapped light spot.This is the concept of Formula(4), and $\omega_{2}(\theta_{T},\theta_{m})$ can be further derived.
		\begin{small} 
			\begin{equation} 
				\frac{\Phi_{T}}{\Phi_{R}}=\frac{R_{T}^{2}}{R_{M}^{2}-R_{T}^{2}}
			\end{equation} 
		\end{small}
		\par $R_M$ is the maximum area radius of optical system illumination.
		\subsection{Design method of compound ellipsoid cavity}
		The composite ellipsoid cavity is formed by discrete superposition of several ellipsoids sharing the first focal point, and the first focal point is set as the focal point of the aspherical lens. The relative parameters of the target ellipsoid are calculated by the mapping relationship of input and output light and the equipartition algorithm of energy mapping. The superposition of the reflection radiation of the composite ellipsoid cavity and the transmitted radiation of the aspherical lens can form uniform illumination.
		\subsubsection{Ellipsoidal cavity energy mapping relationship}
		For the distribution of light intensity with Angle, the total radiation flux $\Phi_{ALL}$ and the radiation flux $\Phi_{i}$ of each inclination range, the reflected radiation flux $\Phi_{R}$ and the transmitted radiation flux $\Phi_{T}$ are given by Formula(5) and Formula(6).
		\begin{small} 
			\begin{equation} 
				\begin{aligned}\Phi_i=2\pi\int_{\theta_i}^{\theta_{i+1}}I(\theta)\sin\theta d\theta\quad\Phi_{All}=2\pi\int_{0}^{\theta_m}I(\theta)\sin\theta d\theta\end{aligned}
			\end{equation} 
		\end{small}
		\begin{small} 
		\begin{equation} 
			\Phi_R=2\pi\int_{\theta_T}^{\theta_m}I(\theta)\sin\theta d\theta\quad\Phi_T=2\pi\int_0^{\theta_T}I(\theta)\sin\theta d\theta 
		\end{equation} 
		\end{small}
		\par $I(\theta)$ is the light intensity distribution function after aspherical lens shaping, and the scale coefficient is defined as $\alpha_{i}$ between $\theta_i$ and $\theta_i+1$. The radiation flux $\theta_i$ and reflected radiant flux $\Phi_{R}$ is equal to:
		\begin{small} 
			\begin{equation} 
				\alpha_i=\frac{\Phi_i}{\Phi_R}\quad i=1,2,\cdots N
			\end{equation} 
		\end{small}
		\par According to the law of conservation of energy, the light rays in the region between $\theta_i$ and $\theta_{i+1}$ are all confined to the annular region between the circle of radius $R_i$ and $R_{i+1}$ of the target receiving surface, as shown in Fig. 3.The light rays in the area between $\theta_i$ and $\theta_{i+1}$ will be mapped to the ring area sandwiched between the circles of radius $R_i$ and $R_{i+1}$ of the target receiving surface. Therefore, the mapping relationship between the energy of the light source and the receiving region of the target plane is established, and the following Formula(7) is established.
		\begin{small} 
			\begin{equation} 
				\begin{aligned}\alpha_i=&\frac{\pi(R_i^2-R_{i+1}^2)}{\pi(R_M^2-R_T^2)}=&\frac{\pi(R_i^2-R_{i+1}^2)}{S_{AII}-S_T}&\end{aligned}
			\end{equation} 
		\end{small}
		\par $S_{AII}=\pi R^{2}_{M}$ represents the total area of the target receiving surface; $S_{T}=\pi R^{2}_{T}$ represents the area where the light rays directly radiate to the target receiving surface area.
		\begin{figure}[htbp]
			\centering
			\includegraphics [scale=0.5]{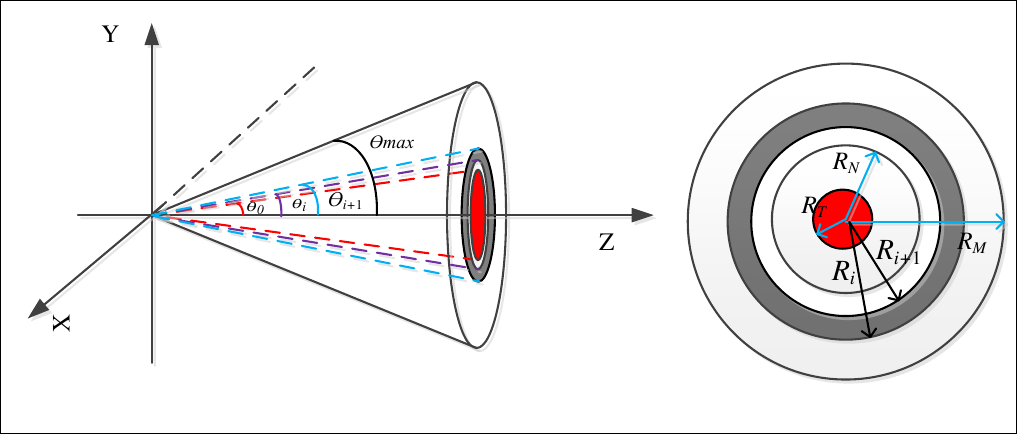}
			\caption{Schematic diagram of luminous flux and receiving plane of light source}\label{fig1}
		\end{figure}
		\par Because the relationship between ring region segmentation of target receiving surface and discretization inclination angle is mapping relationship, therefore, the ring width is set by the Gaus-like beam peak half-height width, which can achieve the superposition uniformity of beam radiation. Because of the nonlinear increasing relationship between the ring area and the radius, through evenly divide the radius of the target receiving surface, so that the realization area of the mapping surface can meet the superposition uniformity. Formula(9) is simplified from Formula(5) to Formula(8), and the corresponding inclination Angle  can be obtained through Formula(9).
		\begin{small} 
			\begin{equation} 
				\int_{\theta_T}^{\theta_{i+1}}I\left(\theta\right)\sin\theta d\theta=\frac{\pi(R_1^2-R_{i+1}^2)\int_{\theta_T}^{\theta_m}I(\theta)\sin\theta d\theta}{S_{R}}
			\end{equation} 
		\end{small}
		\par $S_{R}=S_{All-}S_{T},\quad R_{I}=R_{M},\quad R_{N+I}=R_{T},\quad R_{i+1}=R_{i}+(R_{M-}R_{T})/N.$
		\subsubsection{The design principle of compound ellipsoid cavity}
		\par According to the mapping relationship between the input and output rays of the ellipsoid cavity, the corresponding elliptic parameters and angles of each ellipsoid and the radiation position of the target surface can be obtained, as shown in Fig. 4. In  coordinate system, the light source is located at the origin, this point is also the first focus coordinate of all ovals $F_1$ in the figure), the second focus is set as $F_2$ coordinate $(0,2c_i)$, and the distance from the target receiving plane to the origin is L. 
		\par The radiation flux in Angle $\theta_i$ and $\theta_{i+1}$ is approximately Gaussian, and its peak value is located at Angle 1/2, that is, Angle $\theta^{\prime}_{i}=(\theta_{i}+\theta_{i+1})/2$, the mapping to the target surface is $R_{i}^{\prime}$, and the distance from the target center is $R_{i}^{\prime}=(R_{i}+R_{i+1})/2$. According to the principle of determining a straight line from two points and the optical properties of the ellipse, the relevant parameters of the ellipse  can be obtained by the angle $\theta_i$ and the parameter $R_i$.
		\begin{figure}[htbp]
			\centering
			\includegraphics [scale=0.7]{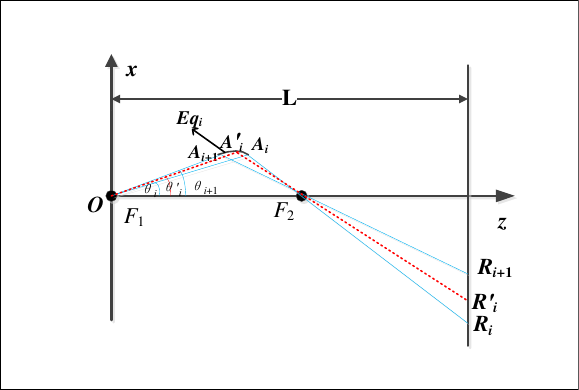}
			\caption{Schematic diagram of complex ellipsoid parametric optical path}\label{fig1}
		\end{figure}
		\par The ellipsoid point $A_{i}^{\prime}(z_{i}^{\prime},x_{i}^{\prime})$ intersecting the Angle rays of $\theta^{\prime}_{i}$ and the straight line determined by the two points $R_{i}^{\prime}$ and $F_2$ must satisfy the elliptic equation, and the elliptic equation $Eq_{i}$ parameter $(a^{\prime}_{i},b^{\prime}_{i},c^{\prime}_{i})$ can be obtained. Thus, the coordinates of the common points of adjacent ellipses $A_{i+1}(z_{i+1},x_{i+1})$ can be further deduced. As the conditions for the parameters of the next elliptic equation $Eq_{i+1}(a^{\prime}{}_{i+1},b^{\prime}{}_{i+1},c^{\prime}{}_{i+1})$, all ellipsoid parameters are obtained through the logical relation in Fig. 5.
		\begin{figure}[htbp]
			\centering
			\includegraphics [scale=0.4]{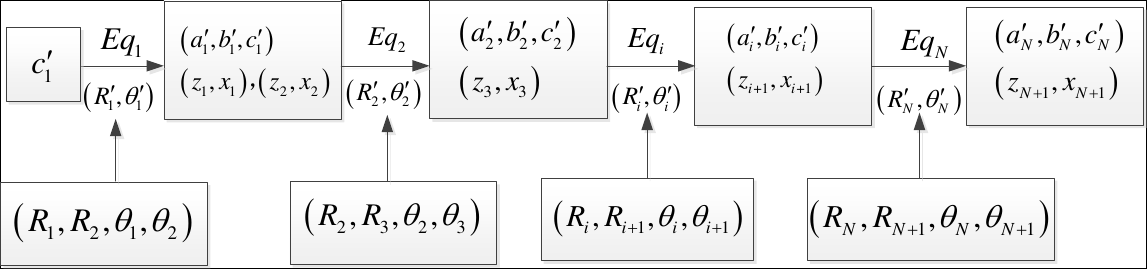}
			\caption{Solving the Logical Relationship Diagram of Composite Ellipsoid Parameters by Energy Mapping Equilibrium}\label{fig1}
		\end{figure}
		\subsubsection{Uniformity optimization scheme}
		If the aspherical focus spot is an extended light source, which is reflected by the composite ellipsoid and projected to the target plane, the energy of the spot edge will decrease. The overlap of the half height and width of the transmission boundary and the inner reflection ring leads to the superposition of energy in the transmission region and the reflection region, and the phenomenon of peak and valley distribution appears. By changing the mapping radius of the inner reflective ring, the superposition region of the reflection transmission is corrected, and the spot uniformity is improved, that is, $R_{N+1}=Ltan(\xi\times\theta_{T})$, where $\xi$ is the regulatory factor. It can be seen from the formula that with the increase of $\xi$, the inner reflective ring moves to the outer reflective ring, and the overlap rate gradually decreases. When the overlap rate is unreasonable, the overlap area presents a peak or trough distribution, as shown in Fig. 6.
		\begin{figure}[htbp]
			\centering
			\includegraphics [scale=0.4]{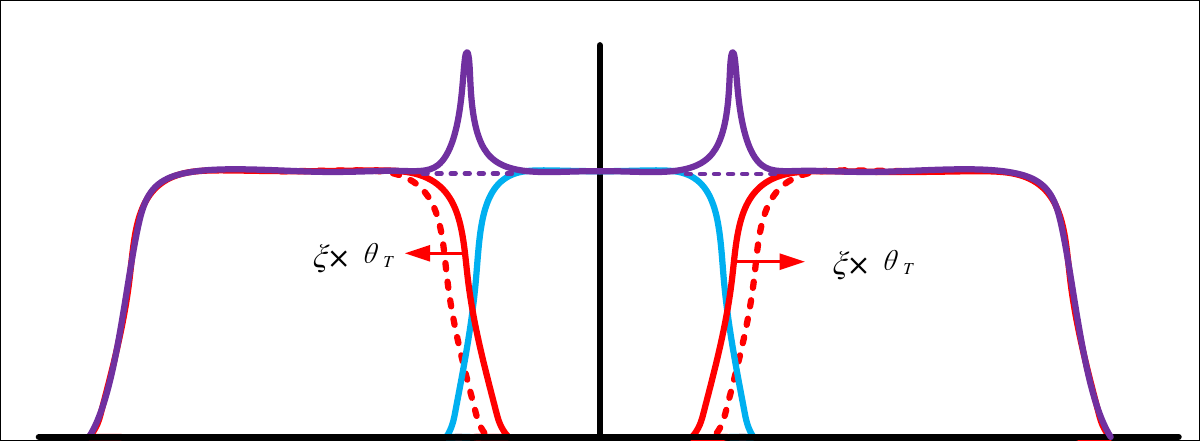}
			\caption{Schematic diagram of energy homogenization}\label{fig1}
		\end{figure}
		\section{Simulation analysis}
		In order to achieve uniform illumination within a distance of $L=2\mathrm{m}$ and a target surface radius of $R_{\mathrm{m}}\mathrm{=1.8m}$, an  laser light source with Gaussian distribution and a waist radius of 5mm is adopted in this paper. The input power $\Phi_{in}$ is 100W and the aperture of aspherical lens is 30mm, with 10mm for thickness. In order to make the length of the system less than 150mm, the semi-focal length $c_{1}{=}60\mathrm{mm}$ of the compound ellipsoidal cavity $L_1$ is set under the aspherical lens.
		\subsection{Simulation analysis of aspherical lens}
		In this paper, the aspherical lens is used to shape the beam so that the beam meets the double Gaussian flat-top distribution. The function relationship $f(r,R,\omega_{1},\theta_{T},\theta_{m})=0$ established by $\mathrm{Formula}(1){\sim}(3)$ indicates that the influence of $R(r)$ on the proportion of reflected transmission radiation will result in the influence of $\theta_{T}$,$\theta_{m}$ on the surface parameters of aspherical lens, thus affecting the size of the focal spot $RMS$. Because the complex ellipsoid cavity is designed for the point light source, the power density uniformity of the radiation reflected to the target plane will be affected if the focused spot is an extended light source. In addition, it can be seen from Formula(4) that the energy of the reshaped light is equal to the area of the mapped light spot, where the ratio of the energy of the flattened part and the energy of the Gaussian part after reshaping is different from the corresponding area ratio, which will have a certain impact on the final uniformity. Therefore, the uniform effect of the reshaped radiation density is evaluated by the effective transmission and reflection power density ratio, which is the product of the actual transmission and reflection power density ratio calculated theoretically, as shown in Formula(610).
		\begin{small} 
			\begin{equation} 
				M_{T/R}=\frac{\Phi’_T}{\Phi’_R}\times\frac{\Phi_R}{\Phi_\tau}
			\end{equation} 
		\end{small}
		\par $\Phi_{T}^{\prime}$ and $\Phi_{R}^{\prime}$ are transmitted and reflected radiation fluxes obtained by optical trace tracing, respectively. When $M_{T/R}$ is 1, it indicates that the design can realize ideal uniform radiation distribution.
		\par By combining $\theta_{T}$ and $\theta_{m}$, the optimization of $M_{T/R}$ and  is realized, and the influence of the introduced aberration of aspherical lens on the complex ellipsoid cavity is reduced, so as to improve the uniformity of the system. Since $\theta_{T}$ and $\theta_{m}$ constrain the composite ellipsoid cavity size and light outlet aperture, $\theta_{m}$ is set in the range of (28,35°) and $\theta_{T}$ is set in the range of (4,8°) for combination matching. The optimization results are shown in Fig. 7 and Fig. 8.
		\begin{figure}[htbp]
			\centering
			\includegraphics [scale=0.8]{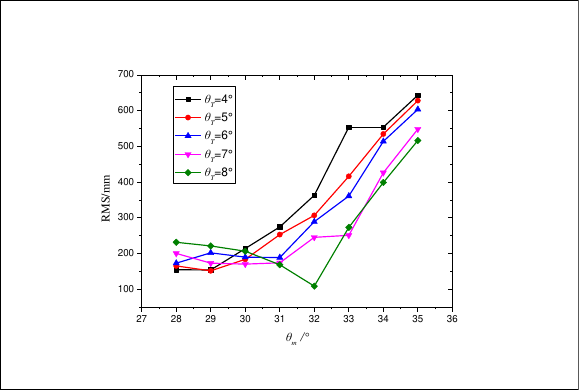}
			\caption{ $M_{T/R}$ changes with $\theta_{T}$ and $\theta_{m}$}\label{fig1}
		\end{figure}
		\begin{figure}[htbp]
		\centering
		\includegraphics [scale=0.8]{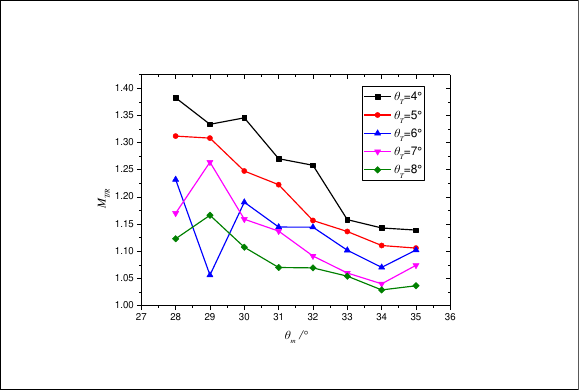}
		\caption{The variation trend of focal spot $RMS$ with $\theta_{T}$ and $\theta_{m}$}\label{fig1}
  		\end{figure}
		\par As can be seen from Fig. 7 and Fig. 8, when the transmission angle $\theta_{T}$ is constant, $M_{T/R}$ will show an overall downward trend while $RMS$ will show an overall increasing trend with the increase of the maximum reflection angle $\theta_{m}$. When the maximum reflection angle $\theta_{m}$ is constant, $M_{T/R}$ and $RMS$ both show a decreasing trend as the transmission angle $\theta_{T}$ increases. According to Fig. 7, the average value of$M_{T/R}$ can be obtained as 1.16. Based on this constraint, it can be concluded that when $\theta_{m}=32^{\circ}$ and $\theta_{T}=8^{\circ}$, the minimum $RMS$ is 108.6um and $M_{T/R}=1.07$ for different combinations $(\theta_{m},\theta_{T})$. The lens parameters are shown in Table 1, and the illumination distribution of the receiving surface cross section after the aspherical lens shaping is shown in Fig. 9.
		\begin{table}[<options>]
			\caption{Parameters related to aspheric lenses.}\label{tbl1}
		\begin{tabular*}{\tblwidth}{@{}LLL@{}}
			\toprule
			\hline
			 & Surface1 & Surface2 \\ [0.5ex] 
			\hline
			Radius/mm & -1.83E-01 & -3.13E+00\\ 
			\hline
			Conic & -1.00E+00 & -6.84E+00\\ 
			\hline
			$r^{2}$ & 2.68E+00 & -7.55E-02\\
			\hline
			$r^{4}$ & -1.23E-03 & 2.23E-04\\
			\hline
			$r^{6}$ & 8.62E-07 & -1.51E-07\\
			\hline
			\bottomrule
		\end{tabular*}
		\end{table}

		\begin{figure}[htbp]
			\centering
			\includegraphics [scale=0.8]{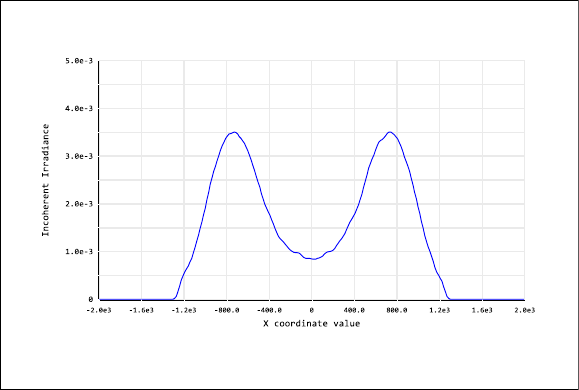}
			\caption{Illuminance distribution across the receiving surface after shaping}\label{fig1}
		\end{figure}
		\subsection{Simulation analysis of complex ellipsoid cavity}
		The light source of the aspherical lens is located at the first focal point of the complex ellipsoid cavity, that is, the origin of the coordinates. The parameters of the complex ellipsoid cavity are solved by mapping uniformity algorithm, and the complex ellipsoid cavity is built by using NURBS surface theory. The optical trace is tracked by Zemax software. Since the overlapping of the energy distribution of the reflected inner ring and the transmitted energy distribution will affect the uniformity effect, the overlap rate is controlled by regulating factors to optimize the uniformity of the target plane. It can be seen from Fig. 10 that $\xi{=}0.9$ has a maximum uniformity of 92.7$\%$. The illumination distribution of the receiving cross section is shown in Fig. 11, and its energy utilization rate is 97.8$\%$.
		\begin{figure}[htbp]
			\centering
			\includegraphics [scale=0.8]{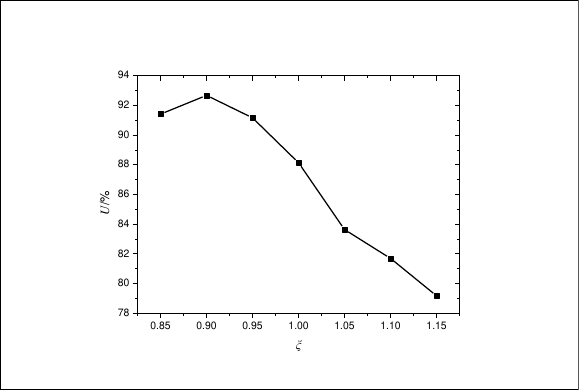}
			\caption{Variation trend of evenness with regulatory factors}\label{fig1}
		\end{figure}
		\begin{figure}[htbp]
		\centering
		\includegraphics [scale=0.8]{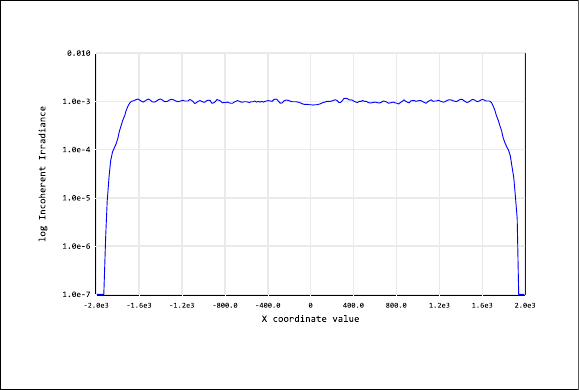}
		\caption{Illuminance distribution of the receiving section}\label{fig1}
	\end{figure}
		\par The composite ellipsoid cavity is composed of each confocal ellipsoid cavity. The axis length and focal length of each ellipsoid cavity $Eq_{i}$ after optimization are obtained by calculation, and the eccentricity of each ellipsoid cavity is obtained according to the axis length and focal length. It can be seen from Fig. 12 that the eccentricity decreases first and then increases with the sequence of $Eq_{i}$ in the ellipsoidal cavity, and its structure presents a conical structure with small sides and large middle. The near infrared lighting system device is designed through aspherical lens and composite ellipsoid cavity. As shown in Fig. 13, this structure has self-adaptive focusing function and installation base, and also has the advantages of small volume and low weight. It is easy to install in the vision system under large field of view, and it can adapt to active lighting scenes under different field of view.
		\begin{figure}[htbp]
			\centering
			\includegraphics [scale=0.8]{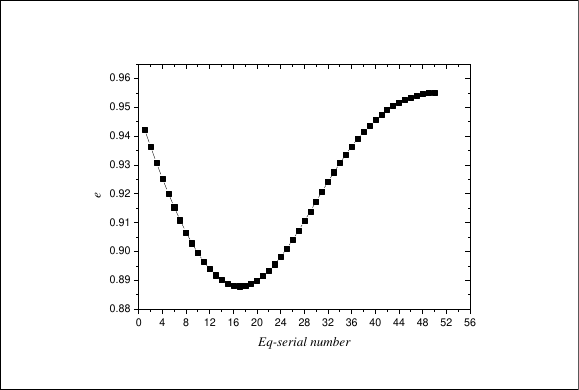}
			\caption{Data diagram of eccentricity of complex ellipsoid cavity}\label{fig1}
		\end{figure}
		\begin{figure}[htbp]
		\centering
		\includegraphics [scale=0.7]{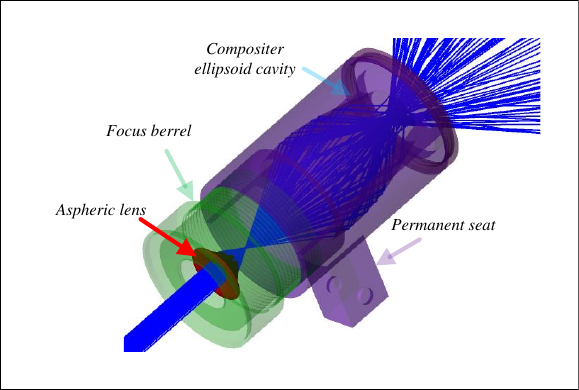}
		\caption{Small portable near infrared lighting device}\label{fig1}
		\end{figure}
		\section{Conclusion}
		In this paper, the uniform illumination of small aperture and large aperture is realized by the combination of an aspheric mirror and a composite ellipsoid cavity. The system is composed of an aspheric mirror and a composite ellipsoid cavity. The aspheric mirror can convert the laser beam with a waist radius of 5mm into a flat-top Gaussian beam. The results show that when the transmission angle is 8°and the reflection angle is 32°, the lens has the best power density and double Gaussian distribution of focal spot $RMS$. The complex ellipsoid cavity is composed of 50 confocal complex ellipsoid cavities, and its parameters are obtained by mapping equipartition theorem. When the regulation factor $\xi{=}0.9$ is set for the confocal composite ellipsoid cavity established by NURBS, a circular spot with a radius of 1.8m can be realized at the position 2m from the target plane of the lighting system. That is, the field of view Angle is 84.0°, the energy utilization rate is 97.8$\%$, the uniformity is 92.7$\%$, and the aperture of the composite ellipsoid cavity is 29.7mm.
		

\begin{thebibliography}{31}
			\bibitem{bib:1}
			HU Haili,et al."Experimental research on imaging of small and dim targets in low illumination background." Infrared and Laser Engineering  49.9 (2020): 20190569-1. (in Chinese)
		\bibitem{bib:2}
			AI Lei. "Overall design method of near infrared laser active night vision lighting system. " Journal of Lighting Engineering 29.3 (2018): 62-65. (in Chinese)
		\bibitem{bib:3}
			Xin-xin Yang. "Laser active illumination imaging quality improvement research. " University of Chinese Academy of Sciences (photovoltaic technology research institute 2021.
		\bibitem{bib:4}
			Luo Wen, et al. "Application of beam shaping in laser illumination. " High Power Laser and Particle Beams 28.7 (2016): 31-35. (in Chinese)
		\bibitem{bib:5}
			Su Yi, et al. "High Energy Laser System" National Defense Industry Press 2004. (in Chinese)
		\bibitem{bib:6}
			ZHANG Lu. "Modeling and Simulation of laser active illumination detection." Aeronautical Weapons 27.04(2020):103-108. (in Chinese)
		\bibitem{bib:7}
			Kumar, Virendra, et al. "Speckle noise reduction strategies in laser-based projection imaging, fluorescence microscopy, and digital holography with uniform illumination, improved image sharpness, and resolution. " Optics $\&$ Laser Technology 141 (2021): 107079.
		\bibitem{bib:8}
			SONG Yanfeng, et al. "Research progress of gate imaging technology for long distance laser illumination. " Laser $\&$ Infrared 43.1 (2013): 9-13. (in Chinese)
		\bibitem{bib:9}
			Molebny, Vasy, et al. "Laser radar: from early history to new trends." Electro-Optical Remote Sensing, Photonic Technologies, and Applications IV. Vol. 7835. SPIE, 2010.
		\bibitem{bib:10}
		 	Nutt, Kyle J., et al. "Developing a portable gas imaging camera using highly tunable active-illumination and computer vision." Optics Express 28.13 (2020): 18566-18576.
		\bibitem{bib:11}
			 Liu Tao, et al. "Design of laser active Lighting optical system." Chinese Optics 9.3 (2016): 342-348. (in Chinese)
		\bibitem{bib:12}
			Zheng, Haijing, et al. "Simulation and Experimental Research on a Beam 		Homogenization System of a Semiconductor Laser." Sensors 22.10 (2022): 3725.
		\bibitem{bib:13}
			Guo Huichao, et al. "Current status of range-gated laser active imaging under atmospheric conditions." Laser $\&$ Optoelectronics Progress 50.10 (2013): 100004-1. (in Chinese)
		\bibitem{bib:14}
			WANG Rui. "Effect of gate width of laser range-gated imaging on signal-to-noise ratio." Chinese Optics 8.6 (2015): 951-956. (in Chinese)
		\bibitem{bib:15}
			 Jiang Hongwang. "Current situation and development of underwater photoelectric detection system." Laser $\&$ Infrared 29.3 (1999): 136-138. (in Chinese)
		\bibitem{bib:16}
			Ma, Yupu,, et al. "Small-divergent-angle uniform illumination with enhanced luminance of transmissive phosphor-converted white laser diode by secondary optics design." Optics and Lasers in Engineering 122 (2019): 14-22.
		\bibitem{bib:17}
			 Zhao Wei, et al. "Design of laser projection display lighting system based on free-form surface array." Acta Optica Sinica 38.6 (2018): 0622001. (in Chinese)
		\bibitem{bib:18}
			 Gorthala, Ravi, et al. "Design and development of a faceted secondary concentrator for a fiber-optic hybrid solar lighting system." Solar Energy 157 (2017): 629-640.
		\bibitem{bib:19}
			Wu, Lei, et al. "Modeling and simulation of range-gated underwater laser imaging systems." International Symposium on Photoelectronic Detection and Imaging 2009: Laser Sensing and Imaging. Vol. 7382. SPIE, 2009.
		\bibitem{bib:20}
			 Wang, Yang, et al. "Design of LED Collimating lens for uniform illumination with freeform surface." 2020 21st International Conference on Electronic Packaging Technology (ICEPT). IEEE, 2020.
		\bibitem{bib:21}
			Madrid-Sánchez, Alejandro, et al. "Freeform optics design method for illumination and laser beam shaping enabled by least squares and surface optimization." Optik 269 (2022): 169941.
		\bibitem{bib:22}
			 Shadalou, Shohreh, et al. "Tunable illumination for LED-based systems using refractive freeform arrays." Optics Express 29.22 (2021): 35755-35764.
		\bibitem{bib:23}
			 Liu, Peng, et al. "Laser regulation for variable color temperature lighting with low energy consumption by microlens arrays." Applied Optics 60.19 (2021): 5652-5661.
		\bibitem{bib:24}
			Chen Daming, et al. "Short-wave infrared imaging zooming laser illumination system." Journal of Infrared and Millimeter Wave 37.3 (2018): 278-283. (in Chinese)
		\bibitem{bib:25}
			Du Yan. "Study on the operating range of near-infrared range-gated imaging System. " Instrument User 19.3 (2012): 4-7.-
		\bibitem{bib:26}
			 Fan, Youchen et al. "Review of development of laser active imaging technology in China and foreign countries." International Symposium on Optoelectronic Technology and Application 2014: Image Processing and Pattern Recognition. Vol. 9301. SPIE, 2014.
		\bibitem{bib:27}
			 Yu, Xunbo, et al. "360-degree tabletop 3D light-field display with ring-shaped viewing range based on aspheric conical lens array." Optics Express 27.19 (2019): 26738-26748.
		\bibitem{bib:28}
			 HAO Jian, et al. "Design of array UV LED uniform lighting system. " Acta Optica Sinica 35.10 (2015): 1022003-1.
		\bibitem{bib:29}
			Kruschwitz, et al. "Laser projection display system." U.S. Patent No. 6,594,090. 15 Jul. 2003.
		\bibitem{bib:30}
			 ZHOU Ping, et al. "Compound eye lens improves energy utilization of liquid crystal projection lighting system " Acta Optica Sinica 24.5 (2004): 587-591. (in Chinese)
		\bibitem{bib:31}
			Niu Lei, et al. " Design of new laser projection display lighting system" Opto-electronic Engineering 42.3 (2015): 77-82. (in Chinese)
		\end{thebibliography}
\end{document}